# All-optical polariton transistor

D. Ballarini[1,2], M. De Giorgi[1,2], E. Cancellieri[3], R. Houdré[4], E. Giacobino[5], R. Cingolani[1], A. Bramati[5], G. Gigli[1,2,6], D. Sanvitto[1,2]

[1]Istituto Italiano di Tecnologia, IIT-Lecce, Via Barsanti, 73010 Lecce, Italy.
[2]NNL, Istituto Nanoscienze - CNR, Via Arnesano, 73100 Lecce, Italy.
[3]Fisica Teorica de la Materia Condensada, Universidad Autonoma de Madrid, Spain.
[4]Institut de Physique de la Matière Condensée, Faculté des Sciences de Base, bâtiment de Physique, Station 3, EPFL, CH-1015 Lausanne, Switzerland
[5]Laboratoire Kastler Brossel, Université Pierre et Marie Curie-Paris 6, École Normale Supérieure et CNRS, UPMC Case 74, 4 place Jussieu, 75005 Paris, France.
[6]Innovation Engineering Department, University of Salento, Via Arnesano, 73100 Lecce, Italy.

**While optical technology provides the best solution for the transmission of information, all-optical devices must satisfy several qualitative criteria to be used as logic elements. In particular, cascadability is difficult to obtain in optical systems, and it is assured only if the output of one stage is in the correct form to drive the input of the next stage. Exciton-polaritons, which are composite particles resulting from the strong coupling between excitons and photons, have recently demonstrated huge non-linearities and unique propagation properties. Here we demonstrate that polariton fluids moving in the plane of the microcavity can operate as input and output of an all-optical transistor, obtaining up to 19 times amplification. Polariton propagation in the plane of the microcavity is then used in turn to control the switching of a second, spatially separated transistor, demonstrating the cascadability of the system. Moreover, the operation of the polariton transistor as an AND/OR gate is shown, validating the connectivity of multiple transistors in the microcavity plane and opening the way to the implementation of polariton integrated circuits.**

Light beams can carry information over long distances, and have proved to be better than electronic wires for low-loss transmissions at high data rate. Moreover, optical connections can, in principle, operate faster and with lower energy consumption than electronic ones even at very short distances, down to chip interconnects [1, 2] with the not negligible advantage of having very low dissipations and virtually no heating. However, the implementation of high-speed, low-energy, optical logics in all-optical integrated circuits represents a big challenge for future information processing [3]. Several systems have been proposed and studied to develop all-optical switches, only to cite some: Mach-Zehnder interferometers in semiconductor materials [4], spin polarization in multiple quantum wells [5], waveguide-coupled ring resonator in silicon [6], photonic crystal nanocavities [7] and polarization bistability in vertical cavity emitting structures [8]. However, to be used as a logic element in an integrated circuit, a switch must satisfy some critical conditions such as cascadability (the output and input should be compatible to allow for connections in series of several devices) and the possibility to feed with one output several inputs (fan-out or gain) and viceversa (fan-in). [9, 10].

In this context, microcavity polaritons, the quantum superposition of electron-hole pairs (excitons) and light (photons), are a peculiar and interesting kind of semiconductor quasi-particles. In fact, polaritons inherit from their photonic part the very light mass ($10^{-4}$ times that of an electron) and high speed (1% of the speed of light), which yields fast switching times [11], while their electronic component provides strong nonlinear effects at power thresholds orders of magnitude lower than in standard dielectrical optical crystals [12-25]. In the last years, the observation of a non-equilibrium condensed phase [26, 27] of polaritons paved the way to the study of new quantum phenomena which could lead to virtually loss-free operations and communication [28-32] up to room temperature [33]. Their potential use for logic operations has also been explored in recent proposals [34, 35], where the information is carried by polariton quasi-particles propagating inside the plane of the microcavity along integrated circuits based on "polariton neurons" [36]. On the other hand, lateral confinement of flowing polaritons has been realized by modelling the photonic potential of the cavity through precise etching of the structure, obtaining propagation distances of more than 200 microns [37].

In this Article we experimentally demonstrate the working principle of a polariton transistor in semiconductor planar microcavities based on the nonlinear interactions between two polariton fluids. The power threshold for nonlinear absorption of a polariton state (Address) is tuned by the injection of a small polariton population (Control), showing a ratio between Address and Control densities of more than one order of magnitude (12dB). We show that the output of one transistor is a polariton fluid propagating in the plane of the microcavity, which is in the correct form to work as a Control signal for the switching *on* of a second, spatially separated, transistor, therefore demonstrating the cascadability of the system. Moreover, the functionality of the polariton transistor is further extended through the implementation of the AND and OR gates, where two propagating polariton fluids are used as inputs for the gate, allowing the proper control of the output level.

## Results

**Experimental configuration.** Our experiments have been performed in transmission configuration and at a temperature of 10 K. A sketch of the experimental configuration is reported in Fig.1a: a single mode, continuous wave, laser beam impinges on a GaAs/AlAs microcavity, composed by front and back Distributed Bragg Reflectors (DBR) with 21 and 24 pairs respectively, and three $In_{0.04}Ga_{0.96}As$ quantum wells embedded in the microcavity plane. The laser beam is partially reflected from the front DBR, which only allows the penetration of those frequencies that match with the polariton dispersion (shown in Fig.1b). The angle of incidence of the laser beam with respect to the direction perpendicular to the sample corresponds to the in-plane momentum of the resonantly created polaritons, which propagates with a finite velocity in the microcavity plane (downward direction in the sketch of Fig.1a). Due to the finite polariton lifetime in the microcavity (around 10ps), this process can be observed through the photon emission from the back DBR: the polariton density in each state is proportional to the emission intensity, which is recorded both in real and momentum space by a CCD coupled to a spectrograph and directly measured by a power meter. In the experiments described in the following, we used linear polarized beams to avoid the effect of the spin on the dynamics of the system. Although the additional degree of freedom constituted by the polariton spin can be advantageous in future devices [38], we will focus here on the demonstration of the basic transistor operation and its connectivity.

**Blueshift mechanism**. An important mechanism which plays a key role in this resonant scheme is the renormalization of the polariton energies with increasing polariton densities. Due to the polariton-polariton interactions, originated from their excitonic components, the polariton resonance is indeed shifted towards higher energies when the polariton density increases. In the case of optical excitation, if the pumping laser is slightly detuned above the polariton resonance, two different regimes can be distinguished. For low enough polariton densities, the laser remains out of resonance (the "off" regime), while for densities above a certain threshold the polariton energy jumps into resonance with the pump energy, resulting in a dramatic increase of the population of the pumped state. This behavior is shown in Fig.1c for the polariton state called Address in this work and described in the following subsection. On the contrary, if the pumping laser is set at resonance with the polariton energy, the polariton density increases linearly until absorption is saturated by the pump-induced blueshift. This is the case for the Control state described below.

**Address and Control state.** In the first part of the experiment we create both Address and Control states by external optical injection, measuring the polariton density in the Control needed to switch *on* the Address. The laser beam has been divided in two paths, which impinge with different angles on the microcavity plane in order to resonantly excite two states of the lower polariton branch, Control and Address, which share the same energy, $E_C = E_A$, but have different finite momenta in the microcavity plane, $K_C$ and $K_A$ respectively. We will see in the second part of the experiment that the control operation can be performed by internal signals, without the need for external laser excitation of the Control state.

In the upper panel of Fig. 1b, the 2-dimensional momentum space $(K_x; K_y)$ image of the emission shows the Control $(K_x = K_C; K_y = 0)$ and Address $(K_x = K_A; K_y = 0)$ states, which are saturated in order to appreciate the weak emission from the microcavity ring due to laser elastic scattering. In the lower panel of Fig. 1b, the polariton dispersion along $K_y = 0$, is shown under non-resonant, low power, excitation. The Control and Address momenta, $K_C = -1 \, \mu m^{-1}$ and $K_A = 0.5 \, \mu m^{-1}$, are indicated by solid vertical lines, while the pumping energy $E = 1.481 \, eV$ is indicated by a dashed horizontal line. We choose this configuration in order to have the detuning $\Delta E_{A,C} = E_{laser} - E_{A,C}$ between the excitation energy $E_{laser}$ and the polariton resonance $E_{A,C}$ for the Address and Control state tuned to $\Delta E_A = 0.4$ meV and $\Delta E_C \approx 0$, respectively.

Due to the slight detuning of the Address state with respect to the laser beam, the polariton population in this state follows the nonlinear behavior shown in Fig.1c, where the emission intensity $I_A$ of the Address state is plotted as a function of the power of the pumping laser. Note that, while the powers of the lasers beams are measured before arriving on the microcavity sample, the amount of light intensity that has penetrated the front microcavity mirror are obtained from the intensities of the Address and Control states, $I_A$ and $I_C$, measured by the photon emission from the back DBR and which correspond to the polariton population in each state. As shown in Fig. 1c, the Address population increases abruptly at the power threshold $P_{th}^A = 5mW$. Increasing further the exciting power, saturation will eventually occur due to the blueshift of the polariton branch with respect to the exciting energy. It is important to note here that the detuning $\Delta E_A$ has been set in a way that any hysteresis cycle (bistable regime) is avoided [18]. Indeed, decreasing the detuning between the exciting energy and the polariton resonance, the width of the hysteresis loop, typical of the optical bistability regime, shrinks towards the presence of only one threshold (optical discriminator regime) [39]. In order to verify that we actually are in this regime, we repeated the measurements of the address intensities under increasing and decreasing excitation powers around the threshold, as shown in Fig. 1d. The black and red dots correspond to points of the onwards and backwards power changes, respectively. As can be seen, the curve is overlapping, showing the absence of the hysteresis power-cycle. This is an important experimental condition to set, in order to obtain a proper control of the switching property.

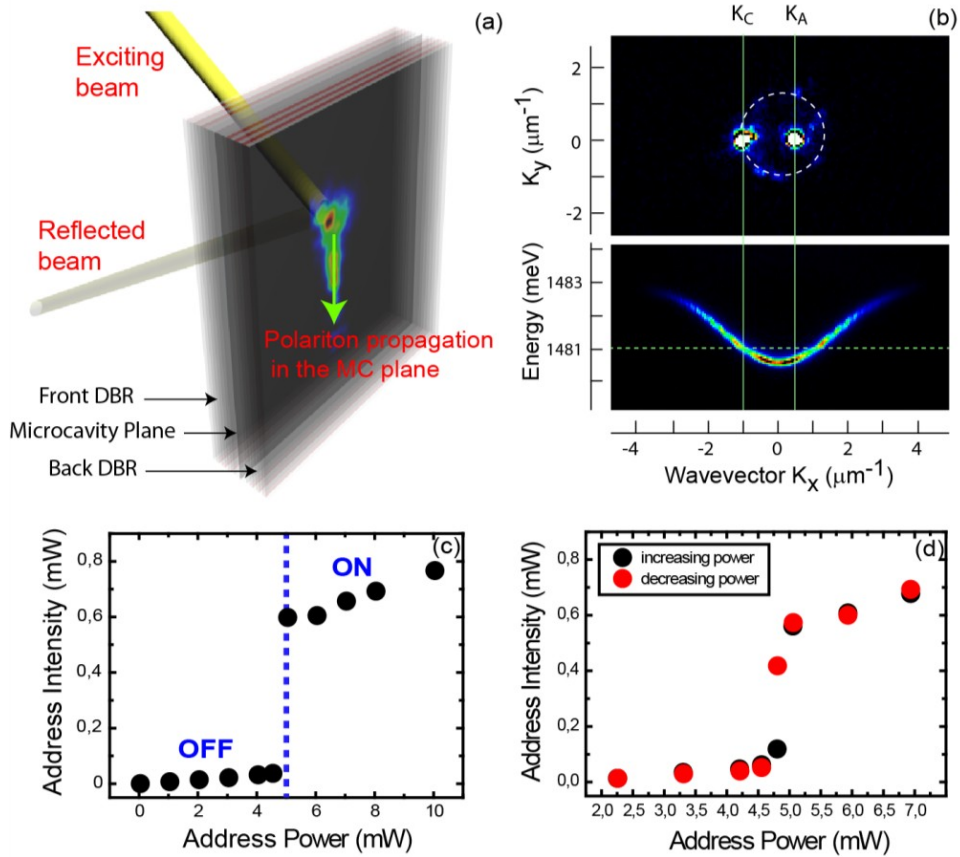

Fig1: (a) Sketch of the experimental configuration: the exciting laser beam is directed with an angle on the microcavity sample and partially reflected from the front DBR. Polaritons are created within the microcavity plane and propagate with a finite velocity. The one-to-one coupling with the external photons allows the observation of the polariton propagation through the emitted intensity from the back side of the sample. (b) In the upper panel, the 2-dimensional momentum space image of the emission under high excitation power shows the polariton states created by two laser beams impinging at different angles ($K_C$ and $K_A$) and same energy. The lower panel shows the polariton dispersion along the direction corresponding to $K_Y=0$ of the upper panel. The vertical lines indicate the value of $K_A=0.5\mu m^{-1}$ and $K_C=-1\mu m^{-1}$ and the horizontal line indicates the pumping energy E=1.481eV. (c) The emission intensities of the address state are plotted for increasing powers of the exciting beam. The Address Powers are measured from the lasers before arriving on the microcavity sample. (d) Measurements of the hysteresis cycle around the threshold of panel (c). Black and red dots correspond to points of the onwards and backwards power changes, respectively, showing the absence of the hysteresis power-cycle.

Conversely, since the Control state is in resonance with the exciting beam, the injection of polaritons occurs with a linear increase of the population even at very low powers, with power-independent insertion losses of 10dB.

The principle of the transistor is to use only a small polariton density in the Control state to switch *on* a much higher polariton density in the Address state. This is indeed possible due to the combination of repulsive interactions between polaritons in different states and the peculiar dispersion given by the light field [40]. Once the Address signal is activated by the external Control, it can be used, in turn, as an internal Control for the next-in-series transistor, in a cascade configuration interconnected through the polariton propagation in the plane of the microcavity, as it will be shown later in this work.

**Transistor Operation**. In Fig. 2a, the characteristic curves of the Address intensity $I_A$ are shown for different values of $I_C$. The threshold power $P_{th}^A$ lowers when increasing the control density, passing from $P_{th}^A = 5\ mW$ in the case of no control beam to $P_{th}^A = 3.5\ mW$ in correspondence of a control intensity of 38 µW. Therefore, for a fixed value of $P^A$, the Address can be switched *on* (green circle in Fig. 2a) and *off* (black circle in Fig. 2a) by changing the polariton density in the Control state.

This operational principle is verified in Fig. 2b, where the Address intensity (blue triangles) is plotted as a function of the Control intensity for $P^A = 3.7\ mW$, corresponding to the red line in Fig. 2a. The address is switched *on* (*off*) for Control intensities $I_C$ above (below) the threshold of 30 µW. The extinction ratio of the Address intensity below and above threshold is $r_e > 10$. The gain $G$ is defined as the ratio between the polariton density in the Address above threshold and the polariton density in the Control state required for the switching operation, and it is directly obtained by measuring the emission intensities of Address and Control at threshold, $G = \frac{I_A}{I_C}$. Note that the gain is defined only by the emission intensities, which are directly related to the number of polaritons inside the cavity. Indeed, the gain requirement consists in the fact that few polaritons in the Control must be able to switch *on* enough polaritons in the Address so that, moving in the plane they can act as a Control for the next transistor in cascade. In this case, a gain $G = 15$ is obtained, showing that the density of the Address is more than one order of magnitude larger than the polariton density in the Control. This threshold corresponds to a density between 5 and 20 polaritons/µm² in the Control state, which means few attojoule/µm². Even considering the relatively wide laser spot used in these experiments to avoid any diffraction effect due to finite cavity width, the polariton population in the whole area is anyhow reaching, at most, an energy of few femtojoule, figure which could be strongly reduced by, for instance, designing *ad hoc* waveguides and mesa structures with lateral size down to a few micrometers.

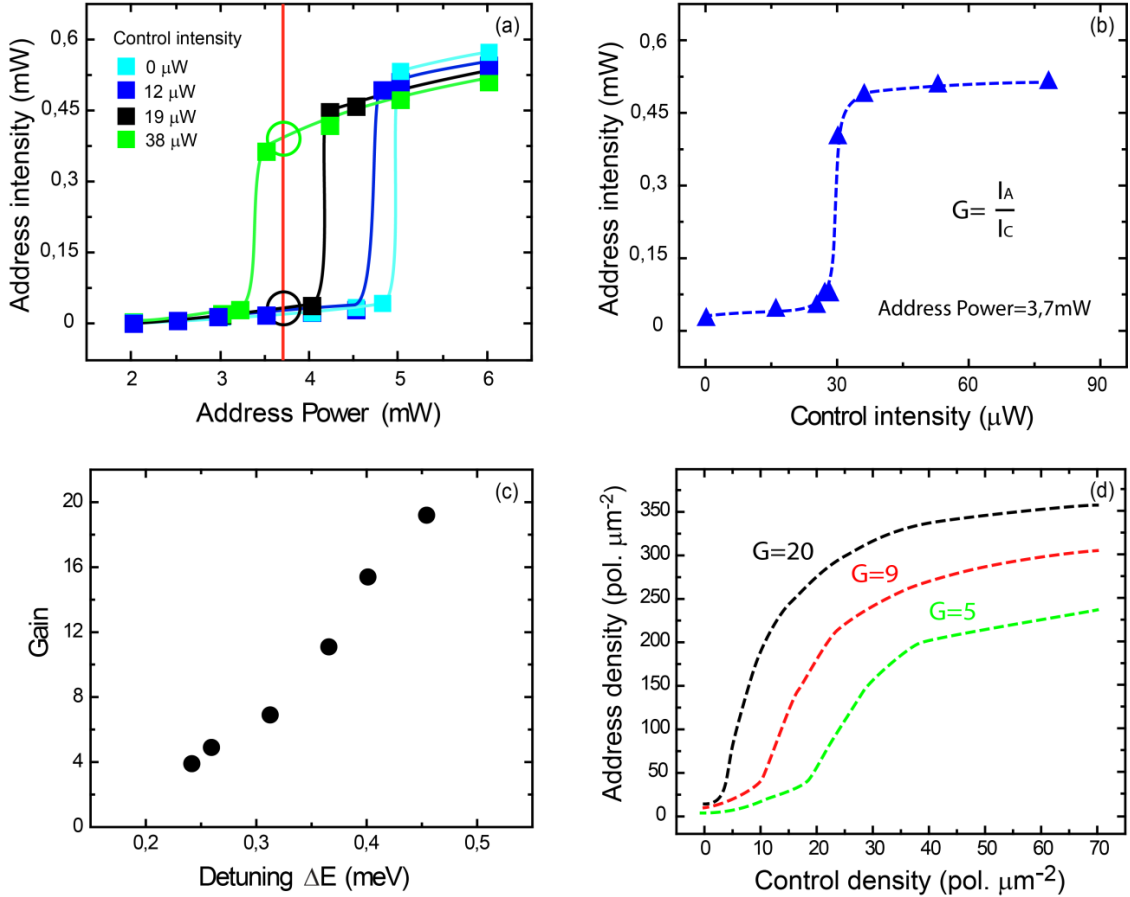

Fig. 2: (a) Address intensity plotted as a function of the Address power for different intensities of the Control: green line ($I_C$=38 μW), black line ($I_C$=19 μW), blue line ($I_C$=12 μW) and without control (light blue line). The red line indicates the power of the Address for which a small change in the Control density brings the Address state from *off* (black circle) to *on* (green circle). (b) Address intensities against Control intensities, taken for an address power of $P^A$=3.7mW, corresponding to the position of the red line in (a). The gain is measured as the ratio between the emission intensities of Address and Control at threshold, $G=I\_A/I\_C$. (c) The amplification ($I_A/I_C$) at threshold is plotted as a function of the energy detuning $\Delta E_A$ between the exciting laser and the polariton branch. The gain increases for higher detuning $\Delta E_A$, showing the possibility to modulate the amount of amplification. (d) Theoretical simulations of the Address densities as a function of the Control density and corresponding gain G for different Address powers, measured in units of the Address threshold $P^A_{th}$: $0.70 P^A_{th}$ (G=5), $0.76 P^A_{th}$ (G=9) and $0.83 P^A_{th}$ (G=20), corresponding to the green, red and black color respectively.

An interesting point is that the value of the amplification at threshold can be modulated by tuning the energy of the exciting laser, obtaining a higher gain when the detuning of the Address $\Delta E_A$ is increased, as shown in Fig.2c. The range of energy detuning $\Delta E_A$ for which the system is still in the optical discriminator regime is wide enough to allow a modulation of the gain from 4 to 19 (note that, in order to obtain a higher gain for larger detuning, the Address power has to be increased accordingly). The polariton densities of Control and Address have been theoretically simulated and confirm the experimental observation of a strong gain at threshold, which is triggered by a polariton density lower than 10 polariton/µm$^2$ in the Control state. These numbers can be improved even further with microcavities of higher finesse and by reducing the Address detuning (though, in this case, with a trade-off of a smaller gain). In Fig.2d, the simulations are run for three different powers of the Address beam (0.70 $P_{th}^A$, 0.76 $P_{th}^A$ and 0.83 $P_{th}^A$, corresponding to the green, red and black color respectively), showing that the control threshold can be lowered by tuning, for instance, the Address power.

**Propagation and Cascadability**. We are now interested in the possibility to demonstrate full cascadability by triggering the absorption of a second Address beam, spatially separated from the first, through the propagation of polaritons within the plane of the microcavity. This scheme, sketched in Fig.3a-b, can be seen as two transistors in series in which the first input controls the second's output. In order to do so, the real and momentum space images of the emission intensities have been simultaneously recorded for the whole area comprising the two transistors. As shown in Fig. 3, the second Address beam (labelled B in Fig. 3) is focused at a distance of 40 µm from the first Address (labelled A), both A and B being below threshold (Fig. 3c and Fig. 3d for real and momentum space respectively). Note that the Control beam (labelled C) is spatially overlapping only with beam A and having orthogonal momentum. In Fig. 3d, the 2D image of the momentum space shows that the momenta of A, B and C in the microcavity plane point in different directions. When the density of C is increased the switching of A results in a polariton fluid moving towards point B, bringing above threshold the second Address B, which in turn propagates in the perpendicular direction. This is shown in Fig. 3e and Fig. 3f, for real and momentum space respectively, with a color scale corresponding to 20 times higher intensity with respect to the *off* regime of Fig. 3c-d. When the Control density in the first transistor is lowered below threshold, both A and B switch *off*.

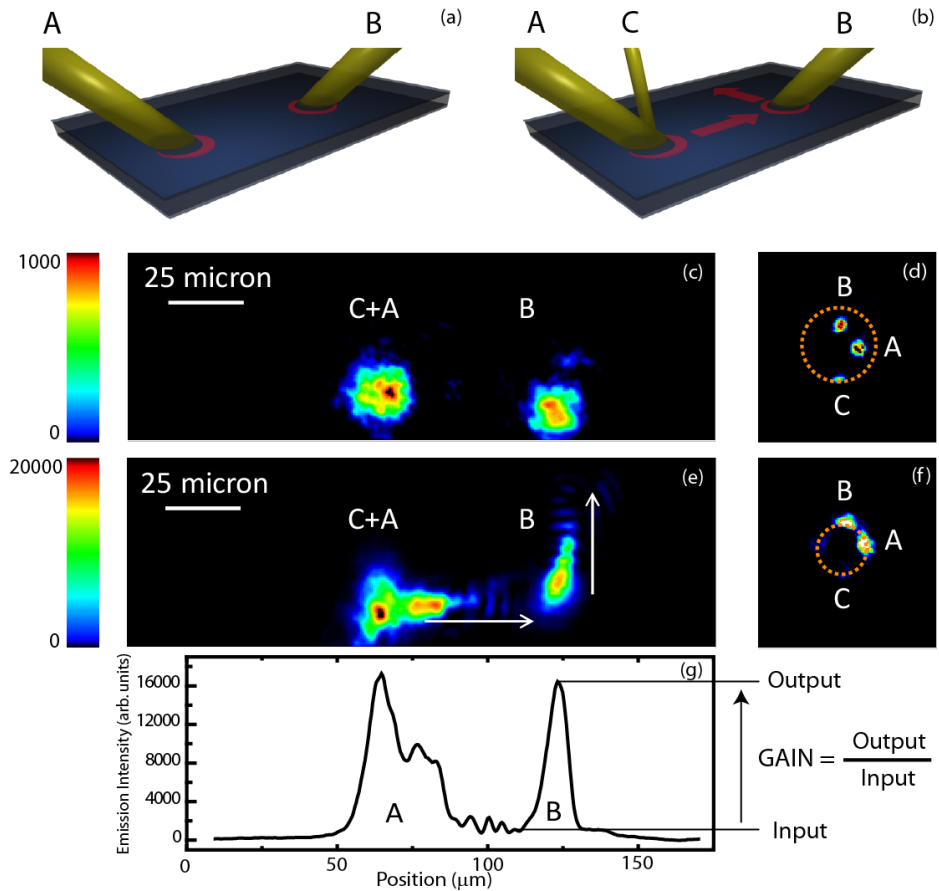

Fig. 3: (a) Scheme of the experimental configuration with two Address beams focused in separated points of the sample with different incident angles. (b) Scheme of the polariton propagation when the Control C is focused over Address A, with the red arrows indicating the propagation direction of the respective polariton states. (c) Real space image of the emission intensities of the Address A and B, and of the Control C below threshold. (d) Far field image of the emission intensities showing the momenta of A, B and C. The circle corresponds to the position of the elastic scattering ring. (e) Real space image of the propagating states A and B, above threshold thanks to the small increase of the intensity of C (same powers as in Fig.2). Note that the intensity in panel (c) has been multiplied 20 times in order to get the same color scale as in panel (e). (f) In the momentum space, A and B are now on resonance with the laser energy and their intensities increase abruptly, while the intensity of C is negligible. (h) Emission intensity profile along the horizontal direction and graphic representation of the gain G.

**AND/OR gates**. Interestingly, two or more Controls can contribute to the activation of the Address, furnishing the basis for the operation of the transistor as a logic gate. We explore here the possibility to use two propagating polariton fluids (labelled A and B in Fig.4) to control the output of a third transistor (labelled C in Fig.4). A and B are separated by a distance of about 70 μm, while C is placed near the intersection point of A and B, as shown in Fig.4, and directed horizontally rightward. The transistor C is indeed working as an AND gate if the threshold is reached when both A and B are *on* (right column of

Fig.4), while it is working as an OR gate if A or B only are required for the activation of C (left column of Fig.4). In Fig.4 the AND gate is obtained by lowering the Address power of C by 10% with respect to the OR case[41]. The upper panel of Fig. 4 shows the OFF state, when both the control of A and B are below threshold.

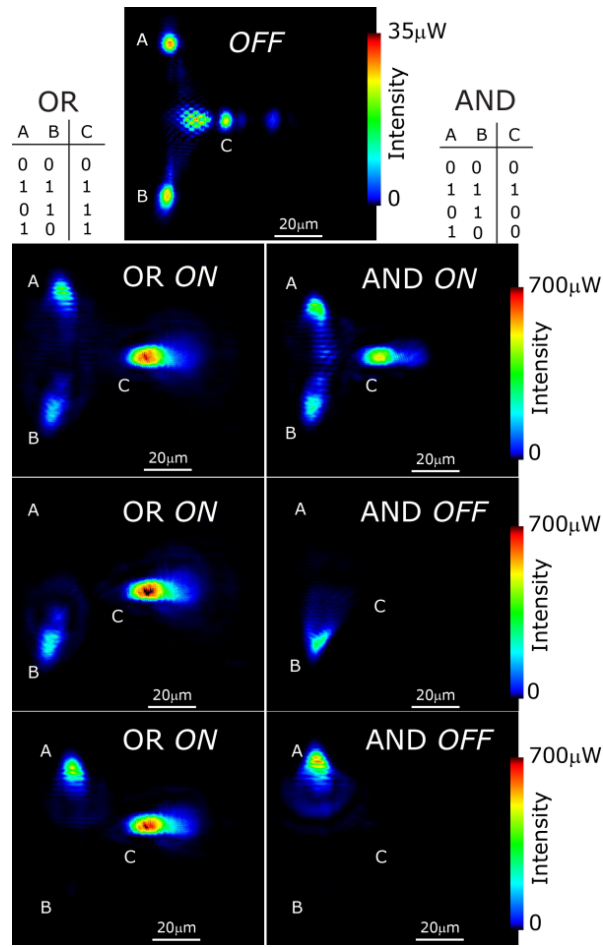

Fig. 4: Three transistors scheme (A,B and C, spatially separated by a distance of about 30 μm) in which C acts as AND/OR logic gate with inputs A and B. When A, B and C are below threshold as in (a), their output intensities are lower than 35 μW. The interference pattern originated by the three fluids is visible near the intersection point in this low density regime. Due to the increase of the intensities above threshold, the color scales in panels (b,c,d,e,f,g) are reduced by a factor 20 as compared with panel (a). The OR gate is described in the left column with both A and B *on* (b), B only *on* (c), A only *on* (d). C is always activated when A or B or both are *on*. The AND gate is obtained by lowering 10% the Address power of C, and it is described in the right column with both A and B *on* (e), B only *on* (f), A only *on* (g). It is required that both A and B are *on* to activate C. The corresponding logic tables are resumed on top of each column.

## Discussion

Optical amplification and switching can be obtained in strongly-coupled exciton-photon systems through the peculiar non-linear properties of microcavity polaritons. The light-matter nature of these quasi-particles allows to obtain switching times of the order of the cavity lifetime (10ps), with the additional advantage of using ultra low activation energies (1 fJ). The results presented in this Article show that using polariton states with finite velocities it is possible to realize an all-optical transistor which is based on the propagation of polariton fluids the chip plane. The on-state propagates out of the injection region towards a second injection point, being able to activate a new polariton fluid with different velocity and direction. Even with the complication to guide polariton fluids in the planar (two-dimensional) microcavity used in this experiment, we demonstrate that one transistor is able to trigger the *on* state of another transistor spatially separated from the first. This operation takes place all inside the microcavity, allowing for multiple interconnections to occur in the plane of the semiconductor chip. We have shown the amplification properties of the transistor in Fig. 2, and proven the cascadability of the system in the experiments of Fig. 3. Indeed, amplification and cascadability are both essential requirements for the connectivity of more transistors in the same circuit. This is further proven in Fig. 4, where two outputs are used as inputs into one transistor, enabling the AND/OR logic operation as a three terminal system (two inputs and one output). Finally, looking at Fig. 2b, it can be observed that the on-state intensity of the output is almost independent from the input intensity (above threshold). This allows the connection of multiple inputs into the same gate, restoring the correct output level.

The results presented so far provide the basis for the integration of polariton systems in future realization of all-optical circuits. In this context, further investigations on the hydrodynamic properties of polaritons should be dedicated to the design of suited geometries for the division and coupling of polariton fluids in guiding structures. In particular, the propagation properties and footprint can be strongly improved using one-dimensional polariton wires etched in high-quality samples (Q > 20000). Note that in this case the finite velocity of the polariton fluids can be controlled independently of the angle of incidence of the injection laser, for example by providing an exciton-photon detuning with a gradient in the desired direction [42] or by controlling the expansion of the polariton fluid [37]. In addition to the classic gates, 0D pillar structures could be also used to exploit the energy quantisation and other phenomena typical of localised superfluids such as the Josephson effects or the conditional tunnelling. This device shows an interesting potential for future optical operation, which could take advantage of the ultra-fast clock cycles and ultra-low energy dissipation of all-optical polaritonic signals.

## Methods

**Two-component Gross-Pitaevskii equation.** A standard way to model the dynamics of the system of resonantly-driven polaritons in a planar microcavity is to use a Gross-Pitaevskii equation for coupled cavity and exciton field ($\psi_C$ and $\psi_X$) generalized to include the effects of the resonant pumping and decay ($\hbar = 1$):

$$i\partial_t = \begin{pmatrix} \psi_X \\ \psi_C \end{pmatrix} = \begin{pmatrix} 0 \\ F \end{pmatrix} + \left[ \hat{H}_0 + \begin{pmatrix} g_X|\psi_X|^2 & 0 \\ 0 & V_C \end{pmatrix} \right] \begin{pmatrix} \psi_X \\ \psi_C \end{pmatrix} \quad (1)$$

where the single particle polariton Hamiltonian $H_0$ reads:

$$\hat{H}_0 = \begin{pmatrix} \omega_X - ik_X & \Omega_R/2 \\ \Omega_R/2 & \omega_C(-i\nabla) - ik_C \end{pmatrix}$$

Where $\omega_C(-i\nabla) = \omega_C(0) - \frac{\nabla^2}{m_C}$ is the cavity dispersion as a function of the in-plane wave vector and where the photon mass is $m_C = 2 \times 10^{-5} m_0$ and $m_0$ is the bare electron mass. For these simulations a flat exciton dispersion relation $\omega_X(k) = \omega_X(0)$ will be assumed and the case of zero detuning at normal incidence $\omega_X(0) = \omega_C(0)$ considered. The parameters $\Omega_R$, $k_X$ and $k_C$ are the Rabi frequency and the excitonic and photonic decay rates respectively and are taken close to experimental values: $\Omega_R = 5.0$ meV, $k_X = 0.001$ meV, and $k_C = 0.1$ meV. In this model polaritons are injected into the cavity in two states (address and control) by two coherent and monochromatic laser fields. The two laser fields share the same smoothen top-hat spatial profile with intensities $F_a$ and $F_c$ for the address and the control respectively, and with full width at half maximum equal to 60 μm. The two laser fields share also the same frequency ω but have different in-plane momentum $k_A = 0.5 \: \mu m^{-1}$ and $k_C = -0.7 \: \mu m^{-1}$. The exciton-exciton interaction strength $g_X$ is set to one by rescaling both the cavity and excitonic fields and the pump intensities. Our theoretical results come from the numerical solution of equation (1) over a two-dimensional grid (256x256) in a 140x140 $\mu m^2$ box using a fifth-order adaptive-step Runge-Kutta algorithm. All the analyzed quantities are taken when the system has reached the steady state condition after 1000 ps. The number of polaritons in the control state at thresholds results to be 3, 5 and 7 polaritons/$\mu m^2$ for the black, red and green curve in Fig. 4d, respectively.

**Evaluation of the polariton densities.** For a measured address emission I = 0.5 mW, we obtain a density $N = \frac{I \times \tau}{E \times A} \approx 80\, pol/\mu m^2$, which corresponds to $16\, aJ/\mu m^2$. The total address energy is $E_{Tot}^A = I \times \tau$, which gives $E_{Tot}^A = 5\, fJ$. Correspondingly, the control density is of about $5\, pol/\mu m^2$, which yields $1\, aJ/\mu m^2$ ($E_{Tot}^C = I \times \tau = 314\, aJ$). In the calculation, we have used $\tau = 10\, ps$ as the polariton lifetime (extracted from the polariton linewidth), while $E = 2 \times 10^{-19}\, J$ is the polariton energy and $A$ is the spot area (radius of $10\, \mu m$). An upper limit can be alternatively obtained from the exciting powers: considering the measured reflection of the address beam from the sample surface (50%), the power of the address is 1.85 mW, and we obtain a density $N = \frac{I \times \tau}{E \times A} \approx 295\, pol/\mu m^2$, which corresponds to $59\, aJ/\mu m^2$. The total address energy is thus $E_{Tot}^A = 18.5\, fJ$. Correspondingly, the control density is of about $20\, pol/\mu m^2$, which yields $4\, aJ/\mu m^2$ ($E_{Tot}^C = 1.3\, fJ$).

**Aknowledgements**

We would like to thank A. Bramanti, V. Scarpa and C. Tejedor for useful discussions, P. Cazzato for the technical assistance with the experiments, and F. M. Marchetti and M. H. Szymanska for the development of the simulation code. This work has been partially funded by the FIRB Italnanonet and the POLATOM ESF Research Networking Program.